# Metric perturbation theory of Quantum Dynamics

Antony L Tambyrajah[1]

*A theory of quantum dynamics based on a discrete structure underlying the space time manifold is developed for single particles. It is shown that at the micro domain the interaction of particles with the underlying discrete structure results in the quantum space time manifold. Regarding the resulting quantum space-time as perturbation from the Lorentz metric it is shown it is possible to discuss the dynamics of particles in the quantum domain.*

Quantum mechanics prescribes the microscopic world to a phenomenal degree of accuracy and has never been found to conflict with experiment. However one of the fundamental issues relating to the interpretation of quantum mechanics is the lack of a coherent theory of dynamics. It has been long recognised that a radical approach to space time may be required,(Monk[1] discusses Wheeler [4] and Schild [2, 3] ) if any meaningful progress is to be made in understanding how particles move in the quantum domain. This question of dynamics is regarded by some as unnecessary in the context of the successful Copenhagen interpretation. Efforts to find new approaches to quantum mechanics are characterised by their incompleteness in the sense that they do not provide a complete and immediate ontological picture of quantum motion .The de Broglie-Bohm

[1] School of Computing and Mathematical Sciences,

The University of Greenwich, Maritime-Greenwich campus, Old Royal Naval College

Park Row, Greenwich, London SE10 9LS

e-mail:ta02@gre.ac.uk

Quantum potential approach, Holland[6], however does provide some ontological explanation of quantum motion. This approach however has its well documented limitations, Bohm [7].

In the approach presented here, a theory of quantum dynamics is developed based on the notion that quantum behaviour is the result of a perturbation of the Lorentz metric. The theory developed here not only addresses some of the key issues relating to the interpretation of quantum motion but will also provide some quantitative results of interest.

The main assertion of the theory is that as a particle moves in the quantum domain it disturbs the space time continuum in its surround which in turn affects the motion of the particle.

The notion that fluctuations in geometry are necessary if one is to explain the quantum behaviour has been suggested for some time. Even before the advent of quantum mechanics there is a pertinent observation by Clifford [8] as discussed in Misner[5]

"I hold in fact

(1) that small portions of space in fact of a nature analogous to little hills on a surface which is on average flat ;namely , the ordinary laws of geometry are not valid in them .

(2) That this property of being curved or distorted is continually passed on from one portion of space to another after the manner of a wave.

(3) That this variation of the curvature of space is what really happens in that phenomenon which we call the motion of matter……. " .

In the context of this paper we can add "this is what determines the motion in the micro domain."

What is presented here is not the usual picture of quantum fluctuations acting as the back drop for quantum motion but rather the transient perturbations in the Lorentz metric as the individual particle travels through space time at the microscopic level.

**Free particles and metric perturbation**

We consider the motion of a free particle moving with constant velocity.

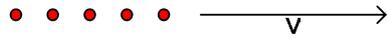

It is postulated that as the particle moves it interacts with the discrete structure underlying space time in such a way as to perturb the Lorentz metric.

We will first outline how we can use, the theoretical formalism of the linearised theory of gravitation, to describe the geodesics resulting from this perturbation (most texts on gravitation, for e.g. Misner[9]). We first write metric as,

$$g_{\mu\nu} = \eta_{\mu\nu} + \psi_{\mu\nu} \tag{1}$$

Where $\eta_{\mu\nu}$ is the Lorentz metric and $\psi_{\mu\nu}$ represents the metric perturbation due to the interaction of the particle with space time background.

With this interaction the resulting possible trajectories of the particle can be determined by the geodesic equation.

$$\nabla_u \nabla_u n + R(...., u, n, u) = 0 \tag{2}$$

In component form the equation is given as

$$\frac{D^2 n^\alpha}{d\tau^2} + R^\alpha_{\beta\delta\gamma} u^\beta n^\delta u^\gamma = 0 \tag{3}$$

Where $R^{\alpha}_{\beta\delta\gamma}$ is the Riemann curvature tensor, $n^{\alpha}$ is the separation vector between two neighbouring geodesics and $u^{\beta}$ is the four velocity of a test particle in the geodesic

The above equation is usually further simplified by introducing the 'proper frame of reference' moving with a particle A and observation is made of particle B under consideration.

In the proper frame the acceleration $a^j=0$ and the rotation of the spatial axes ω attached to the particle A is 0. The observer four velocity in this frame has the components $u_o = -1$ and $u_j = 0$

This reduces the equation to

$$\frac{D^2 n^{\hat{j}}}{d\tau^2} + R^{\hat{j}}_{\hat{0}\hat{k}\hat{0}} n^{\hat{k}} = 0 \tag{4}$$

As A is at the origin, $x_A = 0$ the separation vector $n^{\hat{j}} = x^{\hat{j}}_B$

The equation now can be written as

$$\frac{d^2 x^{\hat{j}}}{d\tau^2} + R^{\hat{j}}_{\hat{0}\hat{k}\hat{0}} x^{\hat{k}} = 0 \tag{5}$$

In the case of plane wave solutions, $\psi_{\mu\nu} = A_{\mu\nu} e^{ik_{\alpha} x^{\alpha}}$, the gauge constraints $A_{\mu\alpha} u^{\alpha} = A_{\mu\alpha} k^{\alpha} = A^{\mu}_{\mu} = 0$ reduces the Riemannian curvature tensor $R^{\hat{j}}_{\hat{0}\hat{k}\hat{0}}$

(There is a TT co-ordinate system that, to a first order in the metric perturbation $\psi^{TT}_{jk}$, moves with particle A and with its proper reference frame. To the first order in $\psi^{TT}_{jk}$ the TT co-ordinate time t is the same as the proper time τ, and $R^{TT}_{j0k0} = R_{j0k0}$) to the Transverse and traceless (TT) form of given as $R^{TT}_{j0k0}$.

Then the equation of motion is

$$\frac{d^2 x^{\hat{j}}}{dt^2} = -R^{TT}_{j0k0} x^{\hat{k}} = \frac{1}{2}(\frac{\partial^2 \psi^{TT}_{jk}}{\partial t^2}) x^{\hat{k}} \qquad (6)$$

Integrating we have

$$x^{\hat{j}}_B(\tau) = x^{\hat{k}}_B [\delta_{jk} + \frac{1}{2}\psi^{TT}_{jk}] + v'^{\hat{j}}_B t \qquad (7)$$

This equation describes the geodesic fluctuations that result and determines the trajectory of the particle. $v'^{\hat{j}}_B$ is the initial velocity of particle at t=0.

A metric perturbation of the Gaussian form,

$$\psi_{jk} = A_{jk} \exp(-k'(z-z')^2/4\sigma) e^{ik'x - i\omega t} \qquad (8)$$

, will be considered in our work. In the last section a possible interaction model for such a perturbation will be discussed based on a discrete space time structure.

The geodesics i.e. trajectory of the particle with the above perturbation will therefore be given as

$$x^{\hat{j}}_B(\tau) = x^{\hat{k}}_B [\delta_{jk} + \frac{1}{2}\psi^{TT}_{jk}] + v'^{\hat{j}}_B t, \qquad (9)$$

Where, $\psi_{jk} = A_{jk} \exp(-k'(z-z')^2/4\sigma) e^{i\omega t}$

The interpretation presented here is that the wave aspect associated with the particle in the quantum domain is in fact the disturbance of the space time continuum caused by the particles.

The intention in this work is to show that this model of the wave particle duality can be used to derive some key results obtained in the conventional quantum mechanics.

**Two slit interference for particles**

The interference fringe pattern produced by particles traveling through two slits is a classic experiment, Jonsson [9], Tonomura [10] demonstrating the wave nature of particles. The results of section 1 showing the generation of geodesic waves will now be used to provide a new theoretical explanation of how the fringe pattern can be produced by the particles.

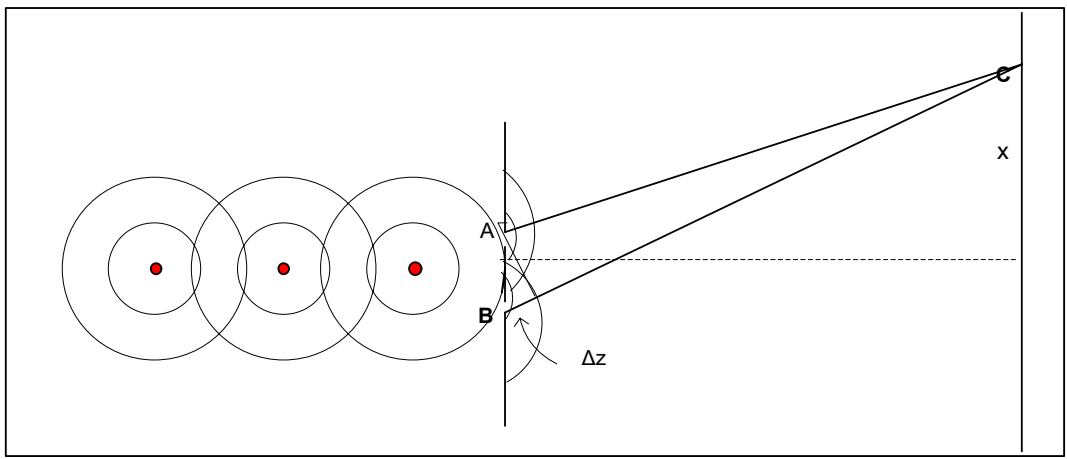

Fig (1) Two slit interference with quantum gravitational waves

The interaction waves generated by the particles as with gravitational waves can be regarded as traveling at the speed of light, although in theory the speed can be even great than the speed of light( Eddington [11]) . Thus the waves are ahead of the particle and reach the slits where they diffract and undergo interference. The transient pattern of geodesic deviation paths formed will determine the path followed by the particles.

The resulting interference pattern is determined by the superposition of waves emanating from slit A and B.

$$\psi^C = \psi^A + \psi^B \tag{10}$$

Where,

$$\psi^A_{jk} = A_{jk} \exp(-k'(z-z')^2/4\sigma)\exp(ik'z - i\omega t)$$
$$\psi^B_{jk} = A_{jk} \exp(-k'(z+\Delta z - z')^2/4\sigma)\exp(ik'(z+\Delta z) - i\omega t) \tag{11}$$

The resulting wave pattern at C is given by

$$\psi^C_{jk} = 2A_{jk}\exp(-k'(z-z')^2/4\sigma)\exp(ik'(z+\Delta z/2) - i\omega t)\cos(k'\Delta z/2)$$

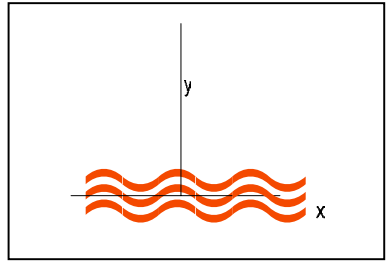

**Fig 2 'quantum gravitational' wave pattern at the screen**

Showing the usual the usual *Cos* dependency along the x direction.

Now according to our proposition this change in the wave pattern represents the new metric perturbation in this region and this in turn will determine the geodesics of the particles in this region.

The metric perturbations in the various directions are

$$\psi^C_{jk} = 2A_{jk}x'\exp(-k'(z-z')^2/4\sigma)\exp(k'(z+\Delta z/2) - i\omega t)\cos(k'\Delta z/2) \tag{12}$$

Now along the z direction at the screen one can consider all the terms in (12) to be of a fixed value except for the last term. We can therefore write (12) as

$$\psi^C_{jk} = A'_{jk}\cos(k'\Delta z/2)\exp i\omega t \tag{13}$$

Where, $A' = 2A_{jk}\exp(-k'(z-z')^2/4\sigma)\exp(ik'(z+\Delta z/2))$

This gives the separation vector for test particle B from the co-moving particle A as

$$x^{\hat{j}}_B(\tau) = x^{\hat{k}}_B[\delta_{jk} + \frac{1}{2}\psi^{TT}_{jk}]$$

$$x_B^{\hat{j}}(\tau) = x_B^{\hat{k}}[\delta_{jk} + \frac{1}{2}A'_{jk}\cos(k'\Delta z/2)\exp i\omega t] \qquad (14)$$

Individual components therefore can be written for the case $A_{31} = A_{32} = A_{33} = 0$

as

$$\begin{aligned}x_B^1 &= x_{t=0}^{\hat{k}}[\delta_{1k} + \frac{1}{2}\psi_{1k}^{TT}] \\ &= x_{t=0}^1 + \frac{1}{2}x_{t=0}^1\psi_{11}^{TT} + \frac{1}{2}x_{t=0}^2\psi_{12}^{TT}\end{aligned} \qquad (15)$$

$$\begin{aligned}x_B^2 &= x_{t=0}^{\hat{k}}[\delta_{2k} + \frac{1}{2}\psi_{2k}^{TT}] \\ &= x_{t=0}^2 + \frac{1}{2}x_{t=0}^1\psi_{21}^{TT} + \frac{1}{2}x_{t=0}^2\psi_{22}^{TT}\end{aligned} \qquad (16)$$

For the sake of clarity we will now label $x_B^1$ as $x$ and $x_B^2$ as $y$. We first consider position of the particles at the slit where $x_{t=0}^1 = x \approx 0$ and therefore we have at the screen from (15)

$$\begin{aligned}x &= \frac{1}{2}y_{t=0}\psi_{12}^{TT} \\ &= \frac{1}{2}y_{t=0}A'_{12}\cos(k'\Delta z/2)\exp i\omega t \\ &= \frac{1}{2}y_{t=0}2A_{12}\exp(-k'(z-z')^2/4\sigma)\cos(k'\Delta z/2) \\ &= A^2{}_{12}\exp(-k'(z-z')^2/4\sigma)\cos(k'\Delta z/2),\end{aligned} \qquad (17)$$

for the case of initial position of particle at the slit in y direction being equal to $A_{12}$.

The amplitude of the oscillation in the x-direction is therefore determined by

$$x = A_{12}^2\cos(k'\Delta z/2) \qquad (18)$$

Where $\Delta z = \dfrac{xd}{D}$

Graphical solution to this is illustrated in Fig. (3).

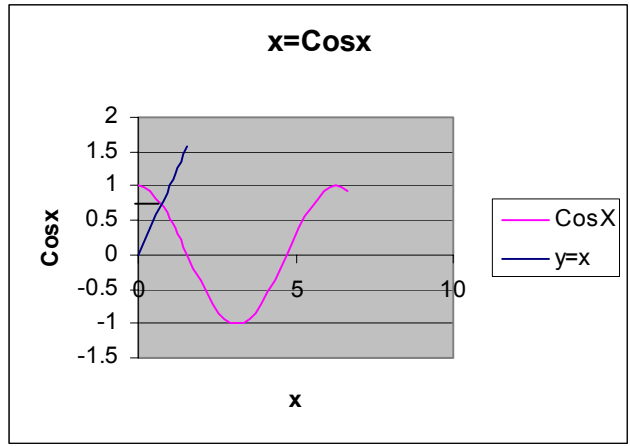

**Fig. (3) Graphical illustration of x=Cosx**

Now if we consider 50kv electrons impinging on slits of separation $0.5\mu$ with a screen at distance 350mm, we obtain width of the fringes.

Data used for numerical of solution of, $x = A_{12}^2 \cos(k'\Delta z / 2)$ with $\Delta z = \dfrac{xd}{D}$

The difference equation in this case is $x_{n+1} = A_{12}^2 \cos(k'\dfrac{d}{D}x_n)$  (19)

**Table I. Data and solution to the difference equation (19)**

$d = 0.5\,\text{E}-11\,\text{m}$,  **D=350 mm**,  $A_{12}^2$ =6E-6,  $\lambda = 5\text{E-}11\,\text{m}$

| | | |
|---|---|---|
| 5.41E-06 | 4.29E-06 | 4.30E-06 |
| 3.39E-06 | 4.31E-06 | 4.30E-06 |
| 4.92E-06 | 4.29E-06 | 4.30E-06 |
| 3.81E-06 | 4.31E-06 | 4.30E-06 |
| 4.65E-06 | 4.30E-06 | 4.30E-06 |
| 4.03E-06 | 4.30E-06 | 4.30E-06 |
| 4.50E-06 | 4.30E-06 | 4.30E-06 |
| 4.15E-06 | 4.30E-06 | 4.30E-06 |
| 4.41E-06 | 4.30E-06 | 4.30E-06 |
| 4.21E-06 | 4.30E-06 | 4.30E-06 |
| 4.36E-06 | 4.30E-06 | 4.30E-06 |
| 4.25E-06 | 4.30E-06 | 4.30E-06 |
| 4.34E-06 | 4.30E-06 | 4.30E-06 |
| 4.27E-06 | 4.30E-06 | 4.30E-06 |
| 4.32E-06 | 4.30E-06 | 4.30E-06 |

This gives the maximum amplitude of the oscillation in the x-direction at the various fringe positions to be 4.03E-6, giving the widths of the fringes obtained to be of the order of $8\mu$.

**Discrete space time**

The suggested quantum space time developed here starts with an underlying lattice structure. Rather than remaining a background structure for the description of quantum theory, it is shown that that the quantum description results from the interaction of particles with the lattice structure of point events. It is further suggested quantum space time obtained it can be used to obtain classical space time by averaging effects of quantum space time.

**Interaction of particles with the discrete space time structure**

The proposed model proposed for discrete lattice is shown in Fig (1)

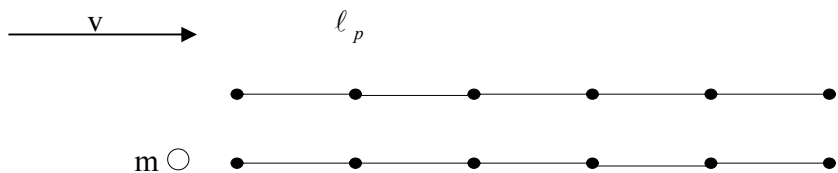
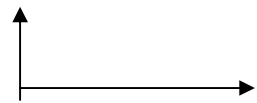

**Fig (4) discrete space time structure**

We first represent these discrete space-time events as pulses, Fig (5)

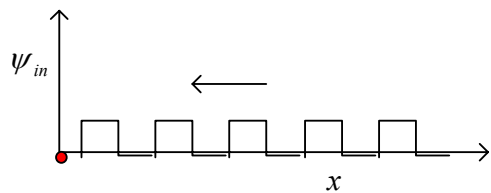

**Fig (5) discrete space structure represented as pulses**

As the particle travels through this discrete space time the particle interacts with these discrete events. The result of the interaction is to convert the discrete digital events into a continuous signal $\psi_{out}$ which can be taken to be the geodesics in the resultant space time manifold in the region of the interaction.

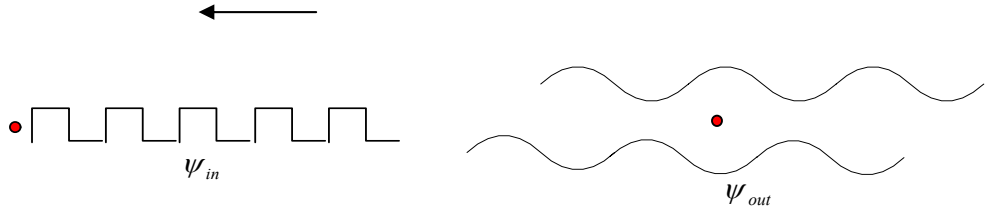

**Fig (6) resultant waves from the interaction of the pulses**

The interaction considered is of the following form, the discrete point events are regarded as a series of pulses $\psi_{in}$ interacting with the incoming particles with characteristic function H.

$\psi_{in}$, can then be represented as

$$\psi_{in} = \delta(x - x')e^{ik'x - i\omega t} \qquad (20)$$

where, $x' = vt$, $v$ is the speed of the particle and

$$\psi_{in} = \int e^{ik(x-x')} dk\, e^{ik'x - i\omega t}, \qquad (21)$$

The output from the interaction $\psi_{out}$, is give by the following equation

$$\psi_{out} = H\psi_{in} \qquad (22)$$

where $H = ae^{i/\hbar \int L dt}$ and $L = p^2/2m$ for a free particle

We have therefore

$$\psi_{out} = \int e^{i/\hbar L \Delta t} e^{ik.(x-x')} dk\, e^{ik'x - i\omega t}, \text{ where } \Delta t = \text{ time interval for the interaction.}$$

$$\psi_{out} = a \int e^{i/\hbar p^2/2m\Delta t} e^{ik.(x-x')} dk e^{ik'x-i\omega t}$$

$$\psi_{out} = a \int e^{-\hbar k^2/i2m\Delta t} e^{ik.(x-x')} dk e^{ik'x-i\omega t} \qquad (22)$$

$$\psi_{out} = a \frac{1}{2\sqrt{\pi\alpha}} e^{-(x-x')^2/4\alpha} e^{iik'x-i\omega t} \quad where \quad \alpha = \hbar \frac{\Delta t}{2mi}$$

This geodesic wave spreads out in both directions causing the particle now to traverse the undulating manifold in an oscillatory motion within the Gaussian envelope.

**Conclusion**

The proposition that particles in the quantum domain disturb the space time continuum in the very local domain has been explored in this paper. The particular Gaussian perturbation was used in the linearised gravitational theory to generate the geodesic equation which in turn determined the motion of the particle. It was shown in section 2 that this model of the quantum motion can be used to explain the two slit interference in the case of particle interference. The question that a purely quantum phenomena can be explained initially in terms of what is essentially an equation that arises from metric considerations should be questionable. However the considerations of section 4 show that the metric perturbation used implicitly assumes the de Broglie wavelength momentum relationship. The motivation for the approach used is that it provides an ontological picture of the particle motion and explains the interference pattern is produced from the succession of particles incident on the slits purely in terms of the motion of the particle. The difficulty associated with explaining how a single particle interferes with itself is avoided in this approach. The last section of the paper contains ideas that are speculative. It assumes that pre quantum space time is not only discrete but consists of point events which behave rather like signals. There is also the suggestion that particles in the micro domain are capable of processing these space-time signals implying an internal structure for the particle which is electronic in nature. These ideas can only be justified if they can be further developed to explain other quantum phenomena such as spin, bound states etc. However in the context of explaining the two slit particle experiment a viable alternative explanation has been presented in this paper.